\documentclass{article}

\usepackage{PRIMEarxiv}

\usepackage[utf8]{inputenc} 
\usepackage[T1]{fontenc}    
\usepackage{hyperref}       
\usepackage{url}            
\usepackage{booktabs}       
\usepackage{amsfonts}       
\usepackage{nicefrac}       
\usepackage{microtype}      
\usepackage{lipsum}
\usepackage{fancyhdr}       
\usepackage{graphicx}       
\graphicspath{{media/}}     
\usepackage{multirow}
\title{Xiaomi-CocktailASR-1 Technical Report
}

\author{
  Xiaoai Plus, ASR team \\
  \\
  Xiaomi Inc., China
}

\begin{document}
\maketitle

\begin{abstract}
Recently, large language model (LLM) based ASR models have achieved significant progress, yet they generally lack support for multi-speaker scenarios, where the cocktail party problem remains a critical bottleneck for further advancing ASR. Existing TS-ASR methods, including end-to-end architectures with speaker embeddings and latest LLM-based explorations suffer from degraded single-speaker performance and the inability to reject when the target speaker is absent. In this paper, we propose Xiaomi-CocktailASR-1, an LLM-based end-to-end TS-ASR architecture. By utilizing reference speech as voiceprint prompts, it directly transcribes the target speaker's speech without requiring speech separation. Xiaomi-CocktailASR-1 maintains competitive performance in single-speaker scenarios, comparable to mainstream ASR models. It also features a negative sample rejection capability, outputting empty text when the target speaker is absent from the mixed speech. Additionally, Xiaomi-CocktailASR-1 supports a Chain-of-Thought (CoT) reasoning mode to provide explicit reasoning steps. Extensive experiments on various synthetic and real-world multi-speaker benchmarks demonstrate that Xiaomi-CocktailASR-1 achieves state-of-the-art performance, effectively addressing the cocktail party problem through a unified architecture that balances multi-speaker and single-speaker recognition accuracy, along with rejection capability.

\end{abstract}


\section{Introduction}
Advances in LLMs have driven mainstream ASR toward Large Audio-Language Models (LALMs) trained on massive datasets with billions of parameters, such as Qwen3-ASR~\cite{Qwen3-ASR}, StepAudio~\cite{step-audio}, and Seed-ASR~\cite{seed-ASR}. 
By fully leveraging the strong modeling capabilities of LLMs, this paradigm integrate various ASR functions into a unified end-to-end architecture, achieving robust performance across multiple languages, dialects, timestamp prediction, and contextual understanding.  
However, existing models struggle with complex multi-speaker scenarios, and no open-source LALMs for TS-ASR have been specifically designed under such cocktail-party settings.

The cocktail party problem is a classic challenge in speech processing, defined as the difficulty of selectively focusing a specific speaker in complex multi-speaker scenarios. Overcoming this problem is essential for the advancement of ASR technology in real-world acoustic conditions. 
Current mainstream solutions for this scenario include multi-talker ASR (MT-ASR) and TS-ASR. 
MT-ASR systems~\cite{MT-ASR-25,MT-ASR-LLM,SOT} transcribe all speakers sequentially, typically following a “first-in, first-out” strategy, with the outputs of different speakers separated by special tokens. However, they cannot associate transcriptions with specific speaker identities, which limits their usefulness when targeting a specific speaker.  
In contrast, TS-ASR~\cite{TS-ASR-2019} leverages reference information from the target speaker to selectively transcribe the target speech while suppressing interference from others, making it more suitable for practical applications.

Traditional TS-ASR research primarily employs cascaded systems combining front-end Target Speaker Extraction (TSE) and back-end ASR, inevitably introducing system complexity and error accumulation~\cite{TS-ASR-conformer, TS-ASR-WavLM}.
Later, some studies explored end-to-end architectures by integrating speaker embeddings with ASR models for joint optimization~\cite{SQ-Whisper, TS-ASR-Whisper, TS-VAD}, but the effectiveness of such approaches is still constrained by independent speaker encoders. 
Recent studies have incorporated LLMs into TS-ASR to leverage their strong semantic capabilities~\cite{MT-LLM,SASE-LLM-TSASR}, yet these approaches still face significant limitations in complex, real-world scenarios.

In real-world interaction scenarios such as smart homes, wearable devices, and intelligent meetings, systems need to not only suppress multi-speaker interference but also dynamically adapt to complex acoustic environments. However, existing TS-ASR systems still face the following limitations in practical applications: 

\begin{itemize}
    \item \textbf{Performance degradation in single-speaker scenarios:} In personal or smart home scenarios, the target user often speaks alone. 
  Existing models designed to suppress multi-speaker interference tend to over-suppress in single-speaker scenarios, which mistakenly suppresses the target speech and increases deletion errors.
  In practical applications, since the number of speakers cannot be predicted in advance, a system designed for either single-speaker ASR or multi-speaker TS-ASR will inevitably underperform in the other scenario. Thus, an ideal model must seamlessly handle both single-speaker and multi-speaker recognition tasks.
  \item \textbf{Lack of rejection capability during target absence:} In outdoor scenarios or meetings, the target speaker may be temporarily absent or silent, which is defined as a negative sample in this paper and requires the model to output empty text.
  However, existing models cannot determine whether the target speaker is present, often transcribing irrelevant speech when the target is absent, which leads to severe false triggers and degrades the user experience.
  \item \textbf{Missing performance and interpretability gains from reasoning:} As a task designed for complex scenarios, TS-ASR is naturally suited for CoT reasoning. 
  Recent studies have shown that CoT reasoning brings performance gains across various speech tasks, and TCP~\cite{think-TS-ASR} specifically demonstrates its effectiveness in TS-ASR. Moreover, the reasoning generated by CoT enhances interpretability and provides valuable information for downstream tasks.
\end{itemize}

\begin{figure*}[t]
\begin{center}
\includegraphics[width=0.9\linewidth]{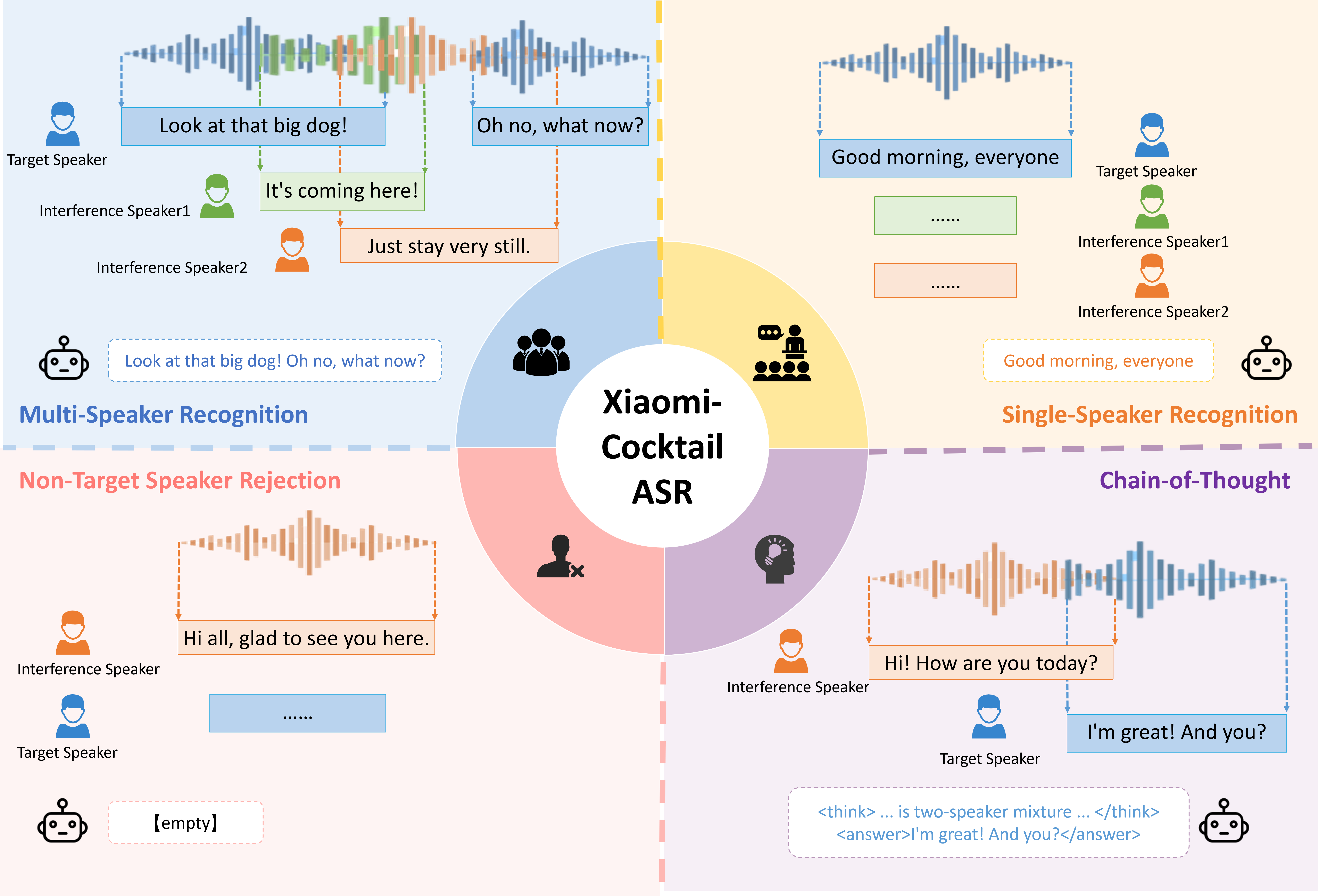}
\end{center}
\caption{The introduction of Xiaomi-CocktailASR-1 capabilities.}
\label{fig:intro}
\end{figure*}

To address these limitations, we propose Xiaomi-CocktailASR-1, an end-to-end TS-ASR system based on LLMs. 
By employing reference speech as voiceprint prompt, the system directly transcribes target speech without relying on explicit speech separation modules. 
As shown in Figure~\ref{fig:intro}, the core contributions of Xiaomi-CocktailASR-1 lie in significantly improving both recognition accuracy and system reliability in complex acoustic scenarios.
Xiaomi-CocktailASR-1 operates in two modes. In the standard mode, a unified prompt enables the model to handle single-speaker recognition, multi-speaker target-speaker ASR, and negative sample rejection within a single architecture. In the CoT mode, the model uses a separate prompt to generate explicit intermediate reasoning steps. Specifically, the main contributions are as follows:

\begin{itemize}
    \item First, it achieves accuracy comparable to mainstream single-speaker ASR models in single-speaker scenarios, eliminating the need to switch models for inputs with different numbers of speakers.
    \item Second, Xiaomi-CocktailASR-1 incorporates a negative sample rejection mechanism that reliably outputs empty text when the target speaker is absent from the input mixture, effectively reducing false triggers in real-world interactions.
    \item Third, Xiaomi-CocktailASR-1 supports CoT reasoning, which generates explicit intermediate logical steps to provide high interpretability for the recognition process.
    \item Finally, the model achieves state-of-the-art performance on multiple simulated and real-world multi-speaker benchmarks.
\end{itemize}

\section{Architecture}

\begin{figure*}[t]
\begin{center}
\includegraphics[width=0.9\linewidth]{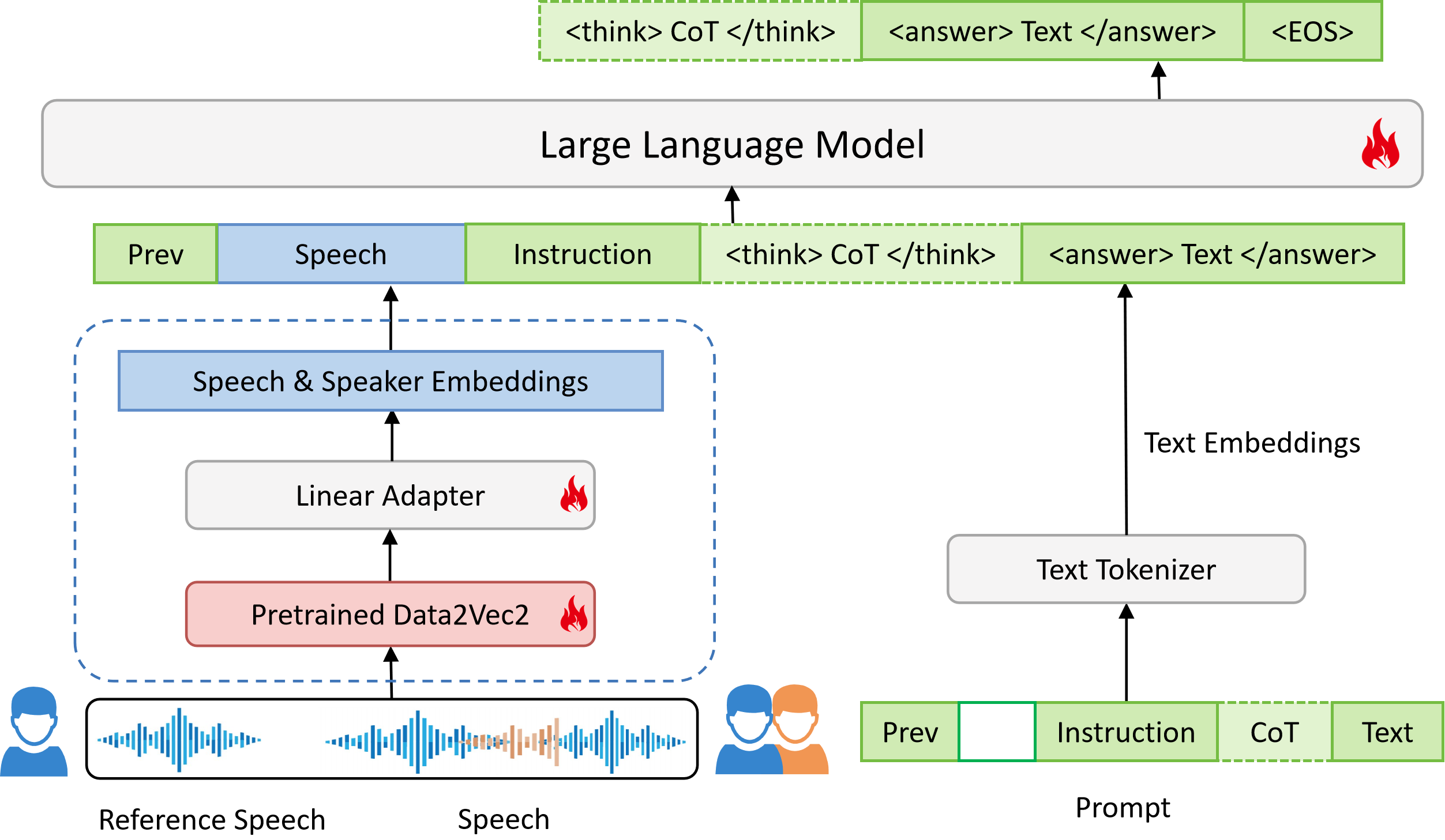}
\end{center}
\caption{The overview of Xiaomi-CocktailASR-1 framework.}
\label{fig:Overview}
\end{figure*}

As illustrated in Figure~\ref{fig:Overview}, the overall architecture of Xiaomi-CocktailASR-1 consists of Audio Encoder, Adapter, and an LLM backbone network. 
The concatenation of reference speech and mixed speech is first fed into the Audio Encoder, which converts it into frame-level embeddings capturing both speech and speaker information, then passed through the Adapter and LLM to generate the target speaker's transcription. 
The model details are described below.

\subsection{Input Organization}

The model input is formed by sequentially concatenating the reference speech, a one-second silence segment, and the mixed speech.
For the reference speech, it is randomly sampled with durations from one to four seconds, providing sufficient information for voiceprint extraction without significant computational cost.
As for the silence segment, it is inserted to explicitly distinguish the reference speech from the mixed signal and anchor the voiceprint features.

\subsection{Model Structure}
As illustrated on Figure~\ref{fig:Overview}, the concatenated audio is first fed into an Audio Encoder with 0.6B parameters.
Derived from the self-supervised learning (SSL) model Data2Vec2~\cite{d2v2}, this encoder converts speech signals into 1280-dimensional frame-level embeddings.
By leveraging the mask prediction mechanism inherent in SSL~\cite{wavlm}, it naturally fuses semantic and speaker information within a single network, thereby eliminating the need for a separate speaker encoder.
Regarding structural improvements, we incorporate a FBank processing module at the front end of the encoder to replace raw waveform inputs. This design significantly accelerates both training and inference speeds while maintaining model performance.
Consequently, when processing multi-speaker mixed audio, the encoder treats the reference audio as a conditional prompt, adaptively focusing on and enhancing the target speaker's speech representations during feature extraction to effectively suppress interference from irrelevant speakers.

The frame-level embeddings output by the Audio Encoder are subsequently processed by an Adapter module.
Employing a lightweight linear network structure, this module performs cross-modal alignment by mapping the embeddings into the hidden feature space of the LLM. 
Simultaneously, a text tokenizer transforms the textual prompt into text embeddings. These aligned audio features and text embeddings are then concatenated and jointly fed into an LLM based on the Qwen3-8B~\cite{Qwen3} base model.
Leveraging the powerful long-sequence modeling and context understanding capabilities of the LLM, the model interprets the joint audio-text input, focuses on the target speaker within mixed speech, and generates high-quality transcriptions of the target speaker.

\section{Training}
To enable Xiaomi-CocktailASR-1 to handle the classic cocktail party scenario as well as various practical situations encountered in real-world applications, the model requires four core capabilities: accurate multi-speaker target speech recognition, single-speaker recognition comparable to standard ASR models, negative sample rejection during target absence, and interpretable CoT reasoning.
To achieve these four objectives, we construct specific training datasets and design a multi-stage pipeline, guiding the model to progressively acquire and balance these capabilities.

\subsection{Mechanism}

\subsubsection{Multi-speaker Recognition}

Multi-speaker recognition is the core capability of a TS-ASR system. 
It requires the model to accurately transcribe the target speech from overlapping multi-speaker speech while effectively suppressing interference from irrelevant speakers.
To achieve this, the model is fine-tuned on large-scale real and synthetic multi-speaker mixtures. 
Furthermore, to enhance robustness in complex acoustic environments, we propose a speech noise mixing strategy during data augmentation. 
By adding environmental noise and background interference with a certain probability, this approach significantly improves the diversity and randomness of the training data. 
Specifically, by probabilistically introducing environmental noise and human speech noise, this strategy generates audio with random overlaps, effectively ensuring the randomness and diversity of overlapping speech in the training data.

\subsubsection{Single-speaker Recognition}

Although TS-ASR systems are designed for multi-speaker scenarios, the number of speakers in practical applications is unpredictable, and single-speaker cases represent the majority. 
For example, the overlap rate in the real-world AliMeeting dataset is merely 30\% to 40\%~\cite{AliMeeting}. 
Therefore, a robust TS-ASR system must not only accurately extract the target speaker from multi-speaker mixtures but also achieve single-speaker recognition performance comparable to standard single-speaker ASR models.
However, the fundamental challenge lies in the inherent trade-off between these capabilities, as enhancing the model's generalization to handle single-speaker scenarios often compromises its specialized performance in suppressing complex multi-speaker interference.

To address this, single-speaker speech data is included in the training set, where reference and target speech pairs are constructed using different utterances from the same speaker. 
By optimizing the sampling ratio of single-speaker and multi-speaker training data, the model effectively mitigates over-suppression, achieving an optimal trade-off between suppressing multi-speaker interference and preserving single-speaker speech integrity.

\subsubsection{Negative Sample Rejection}

Negative sample rejection is defined as the capability to actively reject recognition and output empty text when the target speaker is absent, leaving only interfering speech in the input, thereby effectively preventing the mis-transcription of irrelevant content.
The main challenge of negative sample training is that allowing empty outputs may compromise the model's recognition of normal speech. 
Specifically, the model may incorrectly generate empty text even when the target speaker is actively speaking.

To address this, a negative sample training strategy is proposed, where inputs are concatenated with mismatched reference speech. 
By progressively increasing the proportion of such negative samples across training stages, the model learns to balance accurate rejection with correct recognition.
Moreover, this rejection capability is directly internalized into the model weights through end-to-end supervised learning, requiring no additional rejection thresholds during inference.

\subsubsection{Chain-of-Thought}

The CoT mode equips the model with the ability to explicitly output intermediate reasoning within thinking tags, followed by the final transcription within answer tags.
Specifically, the reasoning includes the number of speakers, as well as each speaker's gender and voiceprint similarity to the reference speech.
This paradigm is realized by training the model on data containing CoT reasoning. 
This explicit reasoning process guides the model to focus on key voiceprint features, effectively reducing recognition errors in complex environments and ultimately improving both accuracy and interpretability.

\subsection{Training Data}

\paragraph{Multi-speaker TS-ASR Data} This dataset comprises three main components, categorized by data source and generation method: (1) open-source multi-speaker data, including real-world overlapping recordings (AMI~\cite{AMI}, AliMeeting~\cite{AliMeeting}) and synthetic mixtures (LibriMix~\cite{librimix}); (2) simulated overlapping multi-speaker data, generated by mixing our proprietary single-speaker ASR data; (3) real-world daily conversational data, collected from both in-house recordings. The total scale of the dataset is approximately 400,000 hours.

\paragraph{Single-speaker TS-ASR Data} In this dataset, both the reference speech and the target speech are derived from different utterances of the same speaker. This setup aims to prevent performance degradation in single-speaker scenarios. The total scale of this dataset is approximately 600,000 hours.

\paragraph{Negative Sample Data} This dataset consists of samples where the reference speaker is absent from the target speech, specifically aiming to train the model's negative sample rejection capability. The total scale is approximately 10,000 hours, with its sampling ratio dynamically adjusted across different training stages.

\begin{figure*}[t]
\begin{center}
\includegraphics[width=1\linewidth]{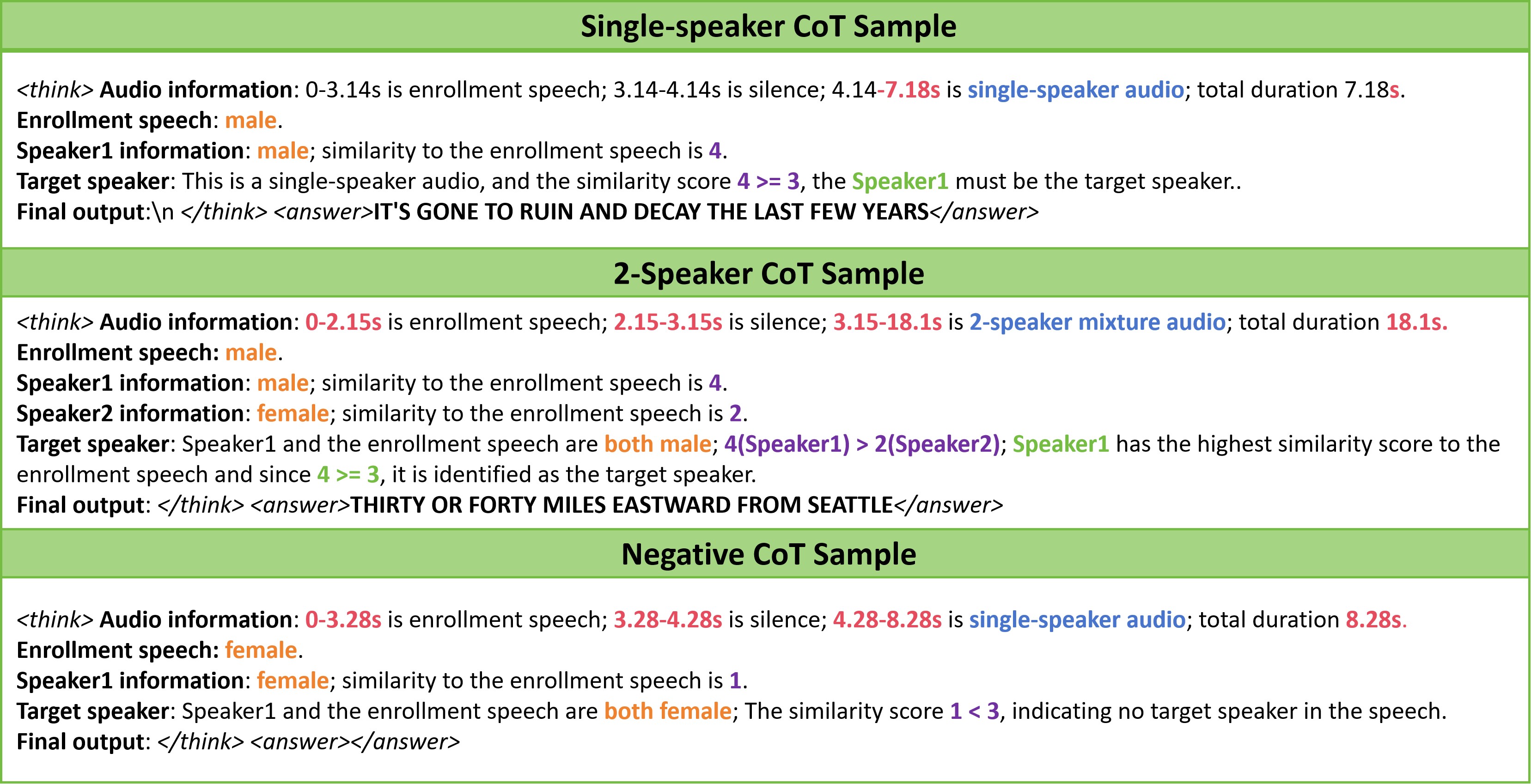}
\end{center}
\caption{Sample of CoT data for single-speaker, 2-speaker mixed speech and negative data. Text in different colors represents different types of information.}
\label{fig:sample}
\end{figure*}

\paragraph{CoT Data} As illustrated in the Figure~\ref{fig:sample}, CoT reasoning are constructed using templates covering multi-speaker positive samples, single-speaker positive samples, and negative samples, each with different formats. 
Template information includes speaker count, gender of each speaker, and similarity levels of each speaker to the reference audio.
Based on this information, the model comprehensively compares gender and speaker similarity to determine whether the similarity falls below a threshold of 3, thereby inferring the identity of the target speaker.
Specifically, the speaker with the highest similarity is identified as the target speaker; if all similarities are below the threshold, it indicates that the target speaker is absent, and empty text should be output.
The speaker with the highest similarity is identified as the target, while if all similarities fall below the threshold, the target speaker is considered absent and empty text is output.

Additionally, similarity levels are computed using the CAM++~\cite{cam++} model to extract speaker embeddings and calculate cosine similarity scores between each source speech and the reference speech embeddings. The continuous similarity scores ranging from 0 to 1 are then mapped to five discrete levels from 1 to 5 through uniform quantization. This discretization prevents the model from focusing on insignificant numerical differences, thereby improving training stability and convergence efficiency.

\subsection{Training Pipeline}

\begin{figure*}[t]
\begin{center}
\includegraphics[width=0.9\linewidth]{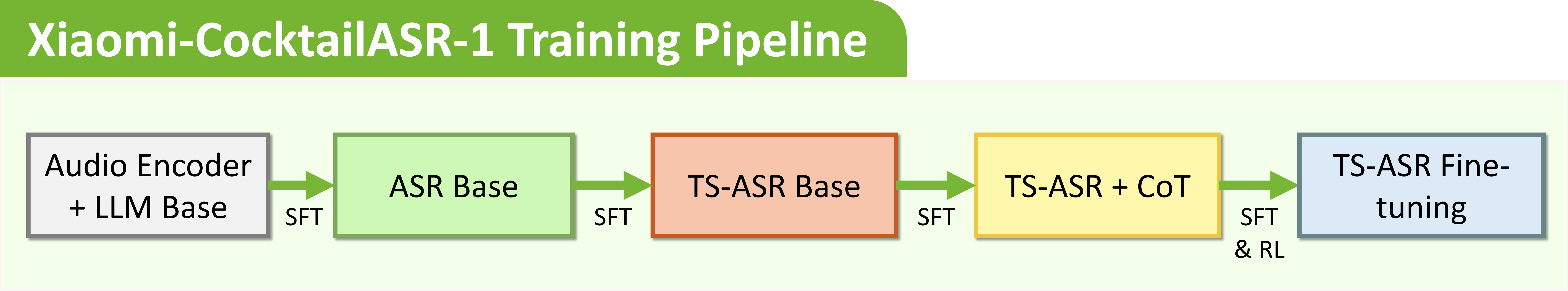}
\end{center}
\caption{The training pipeline of Xiaomi-CocktailASR-1.}
\label{fig:pipeline}
\end{figure*}

\paragraph{Stage 1 ASR Base Model Training} This stage integrates the pretrained Speech Encoder and the LLM base model through an Adapter to achieve cross-modal alignment, enabling the text-focused LLM to adapt to speech recognition tasks.
The model is trained exclusively on standard ASR prompts and regular single-speaker ASR data without reference speech. This establishes a standard ASR Base model, equipping it with basic speech-to-text capabilities.

\paragraph{Stage 2 TS-ASR Base Model Training} Building upon the ASR base model, this stage performs supervised fine-tuning using TS-ASR data and specialized TS-ASR prompts. 
This data is composed of single-speaker and multi-speaker signals, along with the corresponding target speaker reference speech. 
This training process enables the model to extract voiceprint features and locate the target speaker, establishing initial target speaker recognition capabilities.
During this stage, we apply the aforementioned noise mixing strategy with a 50\% probability to directly convert half of the existing data into multi-speaker overlapping speech. 
Furthermore, a small proportion of negative samples is introduced to equip the model with preliminary rejection capabilities without compromising core recognition performance.

\paragraph{Stage 3 TS-ASR CoT Training} This stage incorporates CoT data to introduce reasoning capabilities. 
Specifically, the CoT prompts and data are mixed equally with the original TS-ASR data for joint training.
Through this joint training process, the model learns to generate explicit intermediate reasoning within thinking tags when using the CoT prompt, while directly outputting the transcription under the standard prompt.
Such a design not only provides interpretable reasoning but also maintains excellent performance in the standard mode, allowing users to flexibly enable or disable CoT reasoning based on practical needs.

\paragraph{Stage 4 TS-ASR Model Fine-tuning} Finally, the model is optimized through both supervised fine-tuning and reinforcement learning (RL)~\cite{Deepseekmath} on carefully selected high-quality data to further enhance its comprehensive performance in complex scenarios.
During this stage, the mixing ratios of different training data categories is controlled to ensure a balanced optimization of core capabilities, including multi-speaker recognition, single-speaker generalization, and negative sample rejection.

\section{Evaluation}

\subsection{Benchmarks}

We adopt three categories of evaluation sets to comprehensively assess the model's recognition performance in both single-speaker and multi-speaker scenarios, along with additional negative samples to evaluate its rejection capability.

\subsubsection{Multi-Talker Datasets}

\paragraph{Synthetic Datasets} For synthetic datasets, every speaker in a mixed utterance is alternately treated as the target speaker during evaluation, ensuring objectivity and fairness of the test results.

\begin{itemize}
    \item \textbf{LibriMix}: This is a standard multi-speaker mixture constructed from the LibriSpeech (test-clean) corpus~\cite{LibriSpeech}, with random signal-to-noise ratios (SNRs) ranging from 0 to 15 dB. This random energy distribution ensures that the model cannot rely on shortcuts such as simply transcribing the loudest speaker, thereby truly evaluating its core ability to distinguish speakers based on voiceprint features.
    \item \textbf{LibriSpeechMix}: In this dataset, the starting times of each speaker are randomized to eliminate order-based cues. This ensures that the model cannot rely on sequential shortcuts, such as always transcribing the first speaker, further proving its ability to focus on acoustic voiceprint features.
\end{itemize}

\paragraph{Real-world Datasets}: These datasets mainly consist of far-field multi-speaker recordings with complex spontaneous overlapping speech. Based on official timestamp annotations, the continuous speech is segmented into independent utterances. This ensures that each evaluation segment contains a complete target speaker utterance, while the number of interfering speakers and acoustic conditions stay unpredictable.
\begin{itemize}
    \item \textbf{AMI-SDM~\cite{AMI}}: A classic real-world English multi-speaker meeting corpus recorded with a single microphone. It contains extensive overlapping speech, heavy far-field reverberation, and realistic background noise, making it a strong benchmark for evaluating the model's robustness under challenging acoustic conditions.
    \item \textbf{AliMeeting-Far~\cite{AliMeeting}}: A real-world Chinese multi-speaker meeting corpus recorded using a far-field 8-channel microphone array (Channel 0 is used for evaluation). This dataset is primarily used to evaluate the model's recognition capability in complex Chinese scenarios.
\end{itemize}

\subsubsection{Singer Speaker Datasets}

Single-speaker datasets are primarily used to evaluate whether the model can maintain high recognition accuracy while avoiding deletion errors caused by over-suppression. 
\begin{itemize}
    \item \textbf{LibriSpeech~\cite{LibriSpeech}}: A standard test set for English single-speaker speech recognition, mainly used to evaluate the model's baseline performance in clean environments.
    \item \textbf{AMI-IHM~\cite{AMI}}: Individual headset microphone recordings derived from the AMI meeting corpus.
    \item \textbf{AliMeeting-Near~\cite{AliMeeting}}: Near-field microphone recordings from the AliMeeting corpus, focusing on real-world Chinese meeting scenarios to evaluate the model's generalization ability under Chinese single-speaker conditions.
\end{itemize}

\subsubsection{Negative Datasets}

Negative test sets are specifically designed to evaluate the model's rejection capability when the input contains only interfering speakers. These evaluations verify whether the model can correctly output empty text instead of forcefully transcribing irrelevant speech or generating hallucinations.
\begin{itemize}
    \item \textbf{LibriSpeech Neg}: This test set uses reference speech and input utterances sourced from LibriSpeech test-clean, with the reference randomly selected from speakers not present in the input.
    \item \textbf{Aishell Neg~\cite{aishell1}}: These test sets focus on negative sample evaluation in Chinese scenarios, primarily testing the model's rejection robustness in cross-lingual environments.
    \item \textbf{Chinese in-house Neg}: We collected negative samples from real-world scenarios to evaluate model rejection performance in practical applications.
\end{itemize}


\subsection{Evaluation Metrics}

\begin{itemize}
    \item \textbf{Target Speaker Word / Character Error Rate (TS-WER / TS-CER) )}: These metrics measure the error rates specifically for the target speaker, with the reference text strictly matching the ground truth transcription. The TS-WER and TS-CER are evaluated on the English and Chinese test sets, respectively.
    \item \textbf{Rejection Rate (RR)}: This metric is computed as the proportion of correctly rejected samples among all negative test samples, reflecting the model's ability to reject non-target speakers. A higher rate indicates that the model can reliably detect the absence of the target speaker and produce an empty output, thus reducing the risk of false triggers.
    \item \textbf{False Rejection Rate (FRR) }: This metric is computed as the proportion of incorrectly rejected samples among all positive test samples, measuring the model's failure rate when the target speaker is actually speaking. A lower value is crucial for practical system deployment, as it indicates minimal degradation to standard recognition performance.
    \item \textbf{Non-empty WER}: Since false rejections can cause significant fluctuations in the overall WER, we recalculate this metric by excluding samples with empty outputs. This adjustment eliminates the interference of rejections on WER statistics, providing a pure evaluation of transcription accuracy for non-empty outputs.
\end{itemize}

\subsection{Results}

\subsubsection{Target Speaker ASR in multi-talker senorial}
This section presents a systematic evaluation of the model's comprehensive performance on multi-speaker test sets, which encompass both synthetic datasets and real-world datasets.
Overall, as demonstrated in Table~\ref{tab:model_comparison} and ~\ref{tab:ami_alimeeting_far}, Xiaomi-CocktailASR-1 achieves the SOTA performance across the evaluated benchmarks.

\begin{table}[htbp]
  \centering
  \caption{Performance comparison on different datasets. Evaluated by WER(\%).}
  \label{tab:model_comparison}
  \begin{tabular}{@{}lcccc@{}} 
    \toprule
    \textbf{Model} & \textbf{\begin{tabular}[c]{@{}c@{}}LibriMix\\ 2mix\end{tabular}} & \textbf{\begin{tabular}[c]{@{}c@{}}LibriMix\\ 3mix\end{tabular}} & \textbf{\begin{tabular}[c]{@{}c@{}}LibriSpeechMix\\ 2mix\end{tabular}} & \textbf{\begin{tabular}[c]{@{}c@{}}LibriSpeechMix\\ 3mix\end{tabular}} \\ 
    \midrule
    \textbf{Xiaomi-CocktailASR-1} & \textbf{4.11}  & 12.29 & \textbf{2.90}  & \textbf{4.91}  \\ 
    Qwen3-ASR-1.7b~\cite{Qwen3-ASR}        & 68.75          & 106.04         & 92.17          & 160.82         \\  
    StepAudio2~\cite{step-audio}        & 71.23          & 121.08         & 92.72          & 164.66         \\
    Gemini-2.5-pro\footnotemark[1]~\cite{gemini}           & 48.41          & 76.10          & 30.69          & 52.34          \\
    TCP~\cite{think-TS-ASR} & 4.84       & 12.23         & -           & -           \\ 
    CONF-TSASR~\cite{TS-ASR-conformer} & -       & -        & 5.40           & 7.60           \\
    MT-LLM~\cite{MT-LLM} & 6.70 & 16.2 & - & - \\
    TS-VAD(460h)~\cite{TS-VAD} & 6.61 & 14.81 & 7.92 & 15.97 \\
    Whisper-SS-TTI~\cite{Whisper-MT+TS-ASR} & 7.97 & 21.97 & - & - \\
    Transformer-SA-ASR~\cite{Transformer-SA-ASR} & - & - & 6.40 & 8.50 \\
    \bottomrule
  \end{tabular}
\end{table}
\footnotetext[1]{Using the TS-ASR prompt with reference speech to transcribe only the target speaker, allowing an empty output when the target is absent.}

Specifically, existing top-tier ASR models such as Qwen3-ASR-1.7b and StepAudio2 lack the ability to recognize multiple speakers. Consequently, they are unable to correctly transcribe highly overlapping synthetic datasets, resulting in WERs between 60\% and 160\%.
On the other hand, general multimodal LLMs such as Gemini-2.5-pro can incorporate reference speech through specific prompts, achieving TS-WERs between 30\% and 80\% on synthetic datasets and showing some potential for target speaker ASR.  However, due to the lack of specialized training, their recognition performance in complex acoustic environments remains limited.
In contrast, previous SOTA target speaker ASR models, such as TCP and CONF-TSASR, are specifically optimized for individual datasets, achieving excellent results under a single data distribution but struggling to generalize across all datasets with a single model.
Benefiting from training on large-scale multi-speaker mixed corpora, the proposed Xiaomi-CocktailASR-1 successfully overcomes the above limitations. Compared to previous SOTA models, Xiaomi-CocktailASR-1 achieves significant performance improvements on synthetic datasets, with the TS-WER on LibriMix 2mix dropping from 4.84\% to 4.11\% (a 15.1\% relative reduction) and on LibriSpeechMix 2mix from 5.40\% to 2.90\% (a 46\% relative reduction).
The experimental results clearly demonstrate that Xiaomi-CocktailASR-1 achieves exceptional performance on standard multi-speaker benchmarks, comprehensively outperforming prior TS-ASR models.

\begin{table}[htbp]
  \centering
  \caption{Performance comparison on real-world datasets. Evaluated by WER(\%).}
  \label{tab:ami_alimeeting_far}
  \begin{tabular}{@{}lcc@{}} 
    \toprule
    \textbf{Model} & \textbf{AMI SDM} & \textbf{AliMeeting Far} \\ 
    \midrule
    \textbf{Xiaomi-CocktailASR-1} & \textbf{21.81} & \textbf{20.63} \\ 
    Qwen3-ASR-1.7b        & 38.18          & 39.64            \\ 
    StepAudio2        & 110.50         & 76.82            \\ 
    Whisper Large-v2~\cite{Whisper-large-v2} & 36.40          & -        \\
    Gemini-2.5-pro           & 52.95          & 56.75            \\
    SQ-Whisper~\cite{SQ-Whisper} & 22.0        & -            \\ 
    MC-TS-ASR~\cite{Ali-multi-test} & -        & 27.50            \\ 
    \bottomrule
  \end{tabular}
\end{table}

As shown in Table~\ref{tab:ami_alimeeting_far}, on real-world datasets, ASR models like Qwen3-ASR-1.7b show improved performance compared to their synthetic data results, partly due to the presence of single-speaker segments that mitigate their degradation in complex overlapping conditions. 
However, their WERs still remains above 30\% on these benchmarks, falling far short of practical requirements. For a fair and comprehensive, the comparison focuses on previous SOTA TS-ASR models.
On the AMI-SDM dataset, the previous best TS-ASR model SQ-Whisper achieves a TS-WER of 22.0\%, while Xiaomi-CocktailASR-1 achieves 21.81\%, slightly surpassing this prior result. On the AliMeeting-Far dataset, the previous best MC-TS-ASR model obtains 27.5\%, whereas Xiaomi-CocktailASR-1 substantially reduces the TS-WER to 20.63\%.
Overall, these experimental results demonstrate that Xiaomi-CocktailASR-1 consistently outperforms existing SOTA models and generalizes effectively across diverse datasets and acoustic scenarios.

\subsubsection{Single Speaker ASR}

\begin{table}[htbp]
  \centering
  \caption{Performance comparison on single-speaker datasets. Evaluated by FRR and Non-empty WER (\%).}
  \label{tab:librispeech_ami_ihm}
  
  \small 
  \resizebox{\textwidth}{!}{%
  \begin{tabular}{lcccccccccc} 
    \toprule
    \multirow{2}{*}{\textbf{Model}} & \multicolumn{2}{c}{\textbf{LibriSpeech}} & \multicolumn{2}{c}{\textbf{AliMeeting-near}} & \multicolumn{2}{c}{\textbf{AMI-ihm}} & \multicolumn{2}{c}{\textbf{WenetSpeech(meeting)}} & \multicolumn{2}{c}{\textbf{CommonVoice(zh)}} \\ 
    \cmidrule(lr){2-3} \cmidrule(lr){4-5} \cmidrule(lr){6-7} \cmidrule(lr){8-9} \cmidrule(lr){10-11}
     & FRR & \begin{tabular}[c]{@{}c@{}}Non-empty\\ WER\end{tabular} & FRR & \begin{tabular}[c]{@{}c@{}}Non-empty\\ WER\end{tabular} & FRR & \begin{tabular}[c]{@{}c@{}}Non-empty\\ WER\end{tabular} & FRR & \begin{tabular}[c]{@{}c@{}}Non-empty\\ WER\end{tabular} & FRR & \begin{tabular}[c]{@{}c@{}}Non-empty\\ WER\end{tabular} \\ 
    \midrule
    \textbf{Xiaomi-CocktailASR-1} & 0.36 & 1.73 & 0.38 & 6.57 & 0.01 & 8.89 & 0 & 5.81 & 0.73 & 4.95 \\ 
    Qwen3-ASR-1.7b   & 0             & 1.87          & 0             & 6.39          & 0             & 10.56         & 0             & 5.84          & 0             & 5.39          \\ 
    StepAudio2       & 0             & 1.58          & 0             & 6.82          & 0             & 37.54         & 0             & 5.46          & 0             & 5.07          \\ 
    Whisper Large-v2 & 0             & 2.70          & -             & -             & 0             & 16.90         & -             & -             & 0             & 26.8          \\ 
    Gemini-2.5-pro   & 21.31         & 6.77          & 14.95         & 16.10         & 24.30         & 21.02         & 0             & 27.58        & 0.003         & 14.01        \\ 
    \bottomrule
  \end{tabular}%
  }
\end{table}

As shown in Table~\ref{tab:librispeech_ami_ihm}, Xiaomi-CocktailASR-1 is compared against current mainstream single-speaker ASR models. 
The results show that Xiaomi-CocktailASR-1 performs comparably to these dedicated models on most benchmarks, while achieving substantial improvements on certain challenging datasets.
Specifically, on the AMI-IHM test set, Xiaomi-CocktailASR-1 achieves a WER of 8.89\%, outperforming the best baseline, Qwen3-ASR-1.7b, which yields 10.56\%.
On the LibriSpeech test set, Xiaomi-CocktailASR-1 achieves a WER of 1.73\%, comparable to Qwen2-ASR-1.7b and StepAudio2. 
Furthermore, on both AliMeeting-Near and WenetSpeech(meeting), Xiaomi-CocktailASR-1 demonstrates highly competitive performance with WERs of 6.57\% and 5.81\% respectively, staying on par with the baseline Qwen3-ASR-1.7b (6.39\% and 5.84\%).


Since Xiaomi-CocktailASR-1 have negative sample rejection capability, it may occasionally result in a small probability of false rejection on positive samples. 
As shown in Table~\ref{tab:librispeech_ami_ihm}, Xiaomi-CocktailASR-1 achieves a low FRR of 0.36\% on the LibriSpeech test set, with only a minor impact on the overall WER. 
In contrast, Gemini-2.5-pro exhibits a strong bias toward rejection, yielding a high FRR of 21.31\% in single-speaker scenarios, which substantially degrades its recognition performance.
This comparison suggests that integrating rejection capability without a carefully balanced training strategy tends to result in false rejection. 
By maintaining a low FRR alongside high transcription accuracy, Xiaomi-CocktailASR-1 achieves an effective balance between preserving recognition performance on target-present samples and correctly rejecting target-absent inputs.

\subsubsection{Negative Rejection}

\begin{table}[htbp]
  \centering
  \caption{Performance on negative test sets. Evaluated by RR(\%) $\uparrow$.}
  \label{tab:neg_datasets}
  \begin{tabular}{@{}lccc@{}} 
    \toprule
    \textbf{Model} & \textbf{LibriSpeech Neg} & \textbf{Aishell Neg} & \textbf{Chinese in-house Neg} \\ 
    \midrule
    \textbf{Xiaomi-CocktailASR-1} & 79.59 & \textbf{75.35} & \textbf{68.54} \\ 
    Qwen3-ASR-1.7b        & 0              & 0              & 0 \\ 
    Gemini-2.5-pro           & \textbf{81.79}          & 64.20          & 54.7 \\ 
    StepAudio2        & 0              & 0              & 0 \\ 
    \bottomrule
  \end{tabular}
\end{table}

This section evaluates the rejection capability of the model when the target speaker is absent. 
Experimental results in Table~\ref{tab:neg_datasets} show that existing mainstream ASR models, such as Qwen3-ASR-1.7b and StepAudio2, completely lack a rejection mechanism, yielding rejection rates of 0\%. 
In contrast, Xiaomi-CocktailASR-1 demonstrates stable rejection performance on both the English LibriSpeech Neg and Chinese Aishell Neg test sets, achieving rejection rates of 79.59\% and 75.35\%, respectively. 
Furthermore, on a real-world recorded Chinese in-house negative dataset, Xiaomi-CocktailASR-1 achieves a rejection rate of 68.54\%, indicating its effectiveness in practical scenarios.
When compared to the Gemini-2.5-pro model, which achieves 81.79\% on LibriSpeech Neg and 64.2\% on Aishell Neg, Xiaomi-CocktailASR-1 shows a slightly lower rejection rate in the English scenario but surpasses Gemini-2.5-pro in the Chinese scenario. 

Gemini-2.5-pro exhibits a strong bias toward rejection during inference, meaning its negative sample metrics establish a high rejection baseline. 
The overall rejection performance of Xiaomi-CocktailASR-1 is highly comparable to this strict baseline, demonstrating that our model exhibits an reliable capability to reject negative samples.
Notably, the negative sample test sets are constructed from the same source data as the synthetic multi-speaker and single-speaker test sets, allowing for a direct and fair quantification of this trade-off under identical acoustic conditions.

\subsubsection{Chain-of-Thought}

\begin{table}[htbp]
  \centering
  \caption{Performance on CoT mode. Evaluated by WER(\%).}
  \label{tab:cot_ablation}
  \begin{tabular}{@{}lcc@{}} 
    \toprule
    \textbf{Test Set} & \textbf{Standard mode} & \textbf{CoT mode} \\ 
    \midrule
    LibriMix 2mix      & 4.11 & 3.87 \\ 
    LibriMix 3mix      & 12.287 & 12.285      \\ 
    LibriSpeechMix 2mix & 2.90 & 2.88 \\ 
    LibriSpeechMix 3mix & 4.91 & 4.81 \\ 
    \bottomrule
  \end{tabular}
\end{table}

This section evaluates the performance of the CoT mechanism on synthetic multi-speaker datasets, showed in Table~\ref{tab:cot_ablation}. 
Compared to the standard mode without CoT reasoning, introducing CoT yields a modest but consistent improvement in overall WER. Specifically, the WER on the LibriMix 2mix dataset shows a clear reduction of 0.24\%.
These results clearly demonstrate that the CoT mechanism effectively assists the model in making more accurate decisions in complex scenarios by guiding it through explicit logical reasoning.
Beyond recognition accuracy, the intermediate reasoning steps generated by CoT provide valuable auxiliary information, such as estimated speaker counts and temporal activity patterns, which can potentially benefit downstream tasks requiring deeper speech understanding. This highlights the extensibility of the CoT mechanism beyond mere transcription.

\section{Conclusion}

This paper proposes Xiaomi-CocktailASR-1, an end-to-end TS-ASR architecture based on LLMs. 
By using reference audio as a voiceprint prompt, Xiaomi-CocktailASR-1 directly transcribes target speech without explicit speech separation, achieving state-of-the-art performance on various synthetic and real-world multi-speaker benchmarks. 
Furthermore, the model is equipped with both negative sample rejection and CoT reasoning capabilities, while maintaining competitive recognition performance in single-speaker scenarios.
These results demonstrate the effectiveness of leveraging LLMs for target-speaker ASR, and suggest promising directions for extending such reasoning-enhanced architectures to broader speech understanding tasks, such as multi-party conversation analysis and smart meeting assistance.


\bibliographystyle{unsrt}  
\bibliography{references}  

\section*{Contributors}

All contributors are listed in alphabetical order by their latest names.

\vspace{0.5cm}

\noindent
\begin{minipage}[t]{0.30\textwidth}
  \centering
  \textbf{Core Contributors} \\
  Lichun Fan \\
  Hang Su \\
  Yiru Zhang \\
  
\end{minipage}
\hfill
\begin{minipage}[t]{0.30\textwidth}
  \centering
  \textbf{Contributors} \\
  Tao Li \\
  Lian Li \\
  Yuquan Liang \\
  Chang Liu \\
  Yifeng Wang \\
  Wenhao Yang \\
  Ying Zeng \\
\end{minipage}
\hfill
\begin{minipage}[t]{0.30\textwidth}
  \centering
  \textbf{Supervisors} \\
  Jian Luan \\
  Heng Qu \\
  Cong Zou
  
\end{minipage}

\end{document}